\documentclass[review]{elsarticle}

\usepackage{hyperref}

\journal{Journal of \LaTeX\ Templates}

\biboptions{authoryear}

\newcommand{\msun}{M_{\odot}}

\newcommand{\eg}{{e.g.\ }}

\newcommand{\etc}{{\it etc.\ }}
\def\las{\mathrel{\hbox{\rlap{\hbox{\lower3pt\hbox{$\sim$}}}\hbox{\raise2pt\hbox
{$<$}}}}}
\def\gas{\mathrel{\hbox{\rlap{\hbox{\lower3pt\hbox{$\sim$}}}\hbox{\raise2pt\hbox
{$>$}}}}}

\begin{document}

\begin{frontmatter}

\title{Meteoroid impacts onto asteroids: a competitor for Yarkovsky and YORP}

\author{Paul A. Wiegert}
\address{Dept. of Physics and Astronomy, The University of Western Ontario, London Ontario CANADA\\ \vspace{0.1in}Submitted to Icarus July 14 2014}

\begin{abstract}
The impact of a meteoroid onto an asteroid transfers linear and
angular momentum to the larger body, which may affect its orbit and
its rotational state. Here we show that the meteoroid environment of
our Solar System can have an effect on small asteroids that is
comparable to the Yarkovsky and Yarkovsky-O'Keefe-Radzievskii-Paddack
(YORP) effects under certain conditions.

The momentum content of the meteoroids themselves is expected to
generate an effect much smaller than that of the Yarkovsky
effect. However, momentum transport by ejecta may increase the net
effective force by two orders of magnitude for impacts into bare rock
surfaces. This result is sensitive to the extrapolation of laboratory
microcratering experiment results to real meteoroid-asteroid
collisions and needs further study. If this extrapolation holds, then
meteoroid impacts are more important to the dynamics of small
asteroids than had previously been considered.

Asteroids orbiting on prograde orbits near the Earth encounter an
anisotropic meteoroid environment, including a population of particles
on retrograde orbits generally accepted to be material from
long-period comets spiralling inwards under Poynting-Robertson
drag. High relative speed (60~km/s) impacts by meteoroids provide a
small effective drag force that decreases asteroid semimajor axes
and which is independent of their rotation pole. This effect may exceed
the instantaneous Yarkovsky drift at sizes near and below one meter.

The rate of reorientation of asteroid spins is also substantially
increased when momentum transport by ejecta is included. This has an
indirect effect on the net Yarkovsky drift, particularly the diurnal
variant, as the sign of the drift it creates depends on its rotational
state. The net drift of an asteroid towards a resonance under the
diurnal Yarkovsky effect can be slowed by more frequent pole
reorientations. This may make the effect of the meteoroid environment
more important than the Yarkovsky effect at sizes even above one meter.

Meteoroid impacts also affect asteroid spins at a level comparable to
that of YORP at sizes smaller than tens of meters. Here the effect
comes primarily from a small number of impacts by centimeter size
particles. We conclude that recent measurements of the YORP effect
have probably not been compromised, because the targets large sizes
and because they are known or likely to be regolith-covered rather
than bare rock. However, the effect of impacts increases sharply with
decreasing size, and will likely become important for asteroids
smaller than a few tens of meters in radius.

\end{abstract}

\begin{keyword}
near-Earth objects; asteroid Itokawa; asteroid 2000 PH5; asteroid
dynamics; asteroids rotation; celestial mechanics; impact processes,
meteorites; meteors
\end{keyword}

\end{frontmatter}


\section{Introduction}

The study of the delivery of meteorites to Earth was much advanced by
the revival of the notion that the uneven re-radiation of incident
sunlight could affect the orbits of small asteroids. Known as the
Yarkovsky effect, this phenomenon results when temperature differences
on an asteroid's surface result in it reradiating energy (and hence
momentum) asymmetrically.  The Yarkovsky effect has been widely
discussed elsewhere (the reader is directed to \cite{rub98} and
\cite{farvokhar98} for excellent reviews). It is of interest here
because it is one of few dynamical effects acting in the main asteroid
belt which create a net trend in the semimajor axis $a$ of an
asteroid's orbit. If such a change in $a$ moves the body into a
mean-motion or other resonance, its orbit may be dramatically changed
as a result. Resonances can eject asteroids from the asteroid belt and
play a key role in the delivery of meteorites to Earth. Thus the
Yarkovsky effect, while itself creating only a small change in
asteroid orbits, is nonetheless crucial in moving meteorite parent
bodies from the asteroid belt to near-Earth space. The importance of
the Yarkovsky effect leads one to consider whether or not other small
effects might have important roles in the evolution of small
asteroids.  Here we consider the effect of momentum transfer via
meteoroid impacts on small asteroids and show that it can compete with
the Yarkovsky effect (and its cousin, the
Yarkovsky-O'Keefe-Radzievskii-Paddack or YORP effect) under certain
conditions.

In section \ref{meteoroidenv} we will introduce the meteoroid
environment near the Earth.  In section \ref{effectivedrag}, the
dynamical effects of meteoroid impacts on small asteroids, and in
particular the role of momentum transport by ejecta, will be discussed
and comparisons drawn with the Yarkovsky effect. Section \ref{YORP}
extends the discussion to the YORP effect, section \ref{misc}
considers radiation pressure and rates of erosion and conclusions
are drawn in section \ref{conclusions}.

\section{Meteoroid environment at Earth} \label{meteoroidenv}

Most of the mass accreted by the Earth is in smaller particles, at
least over short times. Larger individual asteroid impacts may
dominate the overall mass input to the Earth \citep{rab93,rabgehsco93}
on million year timescales but they are not relevant here.
\cite{lovbro93} determined that meteoroids with mass $m \approx 1.5
\times 10^{-8}$~kg corresponding to a radius $r=220\mu$m at a density
$\rho_p = 2500$~kg~m$^{-3}$ dominate the meteoroid flux at
Earth. Earlier studies such as those of \cite{gruzoofec85} found
similar values though with total fluxes somewhat (2-3 times) lower.

At these sizes, the meteoroid environment of the Earth is
asymmetric. This is partly because of the Earth's motion around the
Sun: our planet tends to get hit more on the leading side than the
trailing side. However the asymmetry also originates in part from a
heterogeneous distribution of particle orbits.  Studies of the {\it
  sporadic} meteors (that is, those meteors distinct from {\it meteor
  showers}) show concentrations of meteoroid orbits towards the
direction of the Earth's motion around the Sun (\eg \cite{sto86,
  brojon95,chawoogal07, cam08} and many others).  When displayed in a
co-moving reference frame centred on the apex of the Earth's way, a
number of concentrations of impinging orbits are discerned. Here we
will be most interested in those known as the north and south apex
sporadic meteor sources.

Meteoroids arriving at Earth from these apex sources have relative
velocities peaking at 60~km/s \citep{jonbro93,chawoogal07}. These
particles are on approximately circular retrograde orbits. Attributed
to long-period and Halley-family cometary debris that has decayed onto
low-eccentricity orbits through Poynting-Robertson drag, these
particles constitute the dominant momentum and kinetic energy flux in
near-Earth space. Because they arrive from the direction of the
Earth's motion, they hit our planet essentially head-on and provide a
small but consistent tangential drag force on any body (such as an
asteroid) on a similar orbit. Though the meteoroid environment at the
asteroid belt is not well known, it is reasonable to assume that it is
similar to that at Earth and will also produce a net drag on
asteroidal bodies.

The fraction of retrograde meteoroids arriving at Earth has been
measured but there are still uncertainties.  Radial scatter meteor
radars (often called ``High Power Large Aperture'' or HPLA radars)
typically see a larger fraction of apex meteors ($>80$\%)
\citep[\eg][]{satnaknis00,hunoppclo04,jannolmei03,chawoo04} while
transverse scatter (or ``meteor patrol'') radars, typically see a
smaller fraction ($\sim 50$\%) \citep[\eg][]{tay95,galbag04} as do
video meteor systems \citep{cambra11}. This effect can be attributed
to the different instrumental sensitivities \citep{wievaucam09} at
different particle sizes and speeds; however here for simplicity we
will assume that the apex meteoroids constitute a fraction $s$=50\% of
the meteoroid population at these sizes. The magnitude of the effect
of this idealized meteoroid environment on a target asteroid will be
calculated first at Earth.

If the meteoroid flux at the Earth is dominated by the apex source as
studies of the sporadic meteors would suggest, then taking the
(cumulative) flux from Fig. 3 of \cite{lovbro93}, where their
differential flux peaks ($m \approx 1.5 \times 10^{-8}$~kg) we get $n
\approx 3 \times 10^{-8}$~m$^{-2}$~s$^{-1}$ where $n$ is the flux of
particles per square meter per second, and $m$ is the particle mass.
Given these conditions, a one-meter radius asteroid on a circular
orbit near the Earth sees roughly three impacts per year, and each of
impactor carries $\sim 10^{-12}$ of the momentum of the
target. We will consider their cumulative effect to be a small effective
drag on the target asteroid.

\section{Effective drag due to meteoroid impacts} \label{effectivedrag}

The impact of a small meteoroid onto an asteroid surface transfers
kinetic energy and momentum to the larger body. Using the impulse
approximation, the force $F$ exerted on the asteroid as a result of a
momentum gain $\Delta p$ during a time $\Delta t$ is $F= \Delta
p/\Delta t$.  The fraction $\eta$ of the incoming momentum received by
the target is unity in the case of a completely inelastic collision,
and could be as high as two in the case of an elastic
collision. However, high-velocity impacts are highly inelastic and we
will adopt $\eta \approx 1$.

The acceleration $f_a = F/M$ imparted to an assumed spherical asteroid
of mass $M$, density $\rho_a$ and radius $R$ being impacted head-on by
the apex meteoroid population as described earlier would be
\begin{equation}
f_a = \frac{s n m v \pi R^2}{\frac{4}{3}\pi \rho_a R^3} = \frac{3 s n m v }{4 R \rho_a} \label{eq:accel}
\end{equation}
where $v$ is the relative velocity.

Lagrange's planetary equations $e.g.$ \cite{roy78} can be used to
calculate the resulting change in semimajor axis $a$ for an asteroid
with zero eccentricity and inclination that is subject to a tangential
acceleration such as that of Eq.~\ref{eq:accel}
\begin{equation}
\dot{a} \approx -\frac{2f_a}{n'} = -\frac{3 s n m v a^{3/2}}{2 \sqrt{G \msun} R \rho_a} \label{eq:adot0}
\end{equation}
where $n'=\sqrt{G \msun/a^3}$ is the asteroid's mean motion.  For an
$R=1$~m target asteroid at 1~AU, the apex meteoroid environment
produces a decrease in semimajor axis of
\begin{equation}
\dot{a} \approx -6.1 \times 10^{-6} \left( \frac{s}{\rm{0.5}} \right) \left( \frac{R}{\rm{1~m}} \right)^{-1} \left( \frac{\rho_a}{3500~\rm{kg~m}^{-3}} \right) ^{-1}    {\rm AU~Myr}^{-1}   \label{eq:adot1}
\end{equation}
This effective drag force is much lower than that of the Yarkovsky effect
in its different variants, by factors of several up to 100 \citep[\eg
  Figure 1 of ][]{farvokhar98}.

Though direct momentum transfer as described above may be negligible
compared to the Yarkovsky effect, there are two subsidiary effects
that may make small asteroids' interactions with the meteoroid
environment important. First, we will show that the ejecta produced by
the impact results in a much larger momentum transfer to the target
than simply that carried by the impactor, magnifying the effective
force.  Secondly, this may also shorten the timescale between
collisional re-orientation of the asteroid's spin axis, an important
consideration for the Yarkovsky effect, particularly the diurnal
variant.

\subsection{Momentum transport by ejecta}

A hypervelocity impact creates a crater on the target, and the amount
of mass removed during this process is often larger than the mass of
the projectile itself. The incoming particle is vapourized on impact
since its kinetic energy content vastly exceeds its internal binding
energy, and the resulting explosive event excavates a crater in the
target.  Consider the impact as seen in the reference frame of the
centre of mass of the impactor-target pair. Taking the impactor's mass
to be $m$ and its impact velocity $v$, a fraction $\epsilon$ of the
impactor's kinetic is converted to kinetic energy of motion of the
ejecta and target, resulting the ejection of a mass $Nm$ of target
material at a velocity $\gamma v$, where $N$ and $\gamma$ are
multiplicative factors that depend on the detailed physics of impact.

After the impact, the ejecta carries away momentum $\gamma N m v$
which by Newton's Third Law is balanced by an opposite momentum
transfer to the target. The ratio of the momentum of the ejecta to
that of the projectile itself we call $\alpha = N\gamma$. If $\alpha >
1$, then the mobilization of ejecta creates an effective force that
exceeds that due simply to the momentum content of the projectile.
Note that $\alpha >1$ does not violate conservation of momentum. The
kinetic energy of the impactor provides energy for the release of
ejecta, and it is conservation of momentum between the ejecta and the
target that provide the drag force that we consider here.

The value of $\alpha$ can be related to $\epsilon$, $N$ and
$\gamma$. Our definition of $\epsilon$ implies that
\begin{equation}
\frac{1}{2}\epsilon m v^2 = \frac{1}{2}Nm(\gamma v)^2 + \frac{1}{2}MV^2 \label{eq:ConsE1}
\end{equation}
where $M$ is the mass of the target after impact, and $V$ its speed. The ratio of
the kinetic energy acquired by the target to that of the ejecta is
\begin{equation}
\frac{\frac{1}{2}MV^2}{\frac{1}{2}Nm(\gamma v)^2} = \frac{MV^2}{Nm(\gamma v)^2}  \label{eq:KEratio}
\end{equation}
Conservation of momentum implies that
\begin{eqnarray}
(Nm)(\gamma v) &=& MV\\
V &=& \frac{\gamma N m v}{M}
\end{eqnarray}
which when substituted back into Eq.~\ref{eq:KEratio} gives
\begin{eqnarray}
\frac{MV^2}{Nm(\gamma v)^2} &=& \frac{M (\frac{\gamma N m v}{M})^2}{Nm(\gamma v)^2}\\
 &=& \frac{Nm}{M}
\end{eqnarray}
The target carries only a fraction $Nm/M$ of the kinetic energy which
is negligible in the limit that the target mass is much larger than
the amount of mass released by the impact.  Since this applies to most
of the impacts we consider here, this allows us to simplify
Eq.~\ref{eq:ConsE1} to
\begin{eqnarray}
\frac{1}{2}\epsilon m v^2 &\approx& \frac{1}{2} Nm (\gamma v)^2\\
\epsilon &\approx & N\gamma^2 \label{eq:mult1}
\end{eqnarray}
From this, we obtain $\gamma \approx \sqrt{\epsilon/N}$ and
\begin{equation}
\alpha = N\gamma \approx \sqrt{\epsilon N} \label{eq:alpha}
\end{equation} 

We will see below that for typical microcratering events expected on
asteroids, $N$ is very large ($\gas 10^4$) and $\alpha \sim 100$,
which pushes the resulting drag force into a range comparable to that
of the Yarkovsky effect for small asteroids. This is a linchpin
argument of this paper, namely that momentum transport by ejecta
creates a substantially larger effective force on the target than if
the simple momentum content of the projectile would imply. Since
hypervelocity cratering is complex, our analysis may be
over-simplified and further study by experts in that field is greatly
encouraged by this author.

\subsection{Microcratering experiments}

Microcratering experiments involve accelerating of particles to high
speeds in the lab and directing them onto targets composed of the
materials of interest. Such experiments often measure the amount of
material excavated $(N)$, while the values for ejecta velocities
($\gamma$) and energies ($\epsilon$) are less well-studied. We will
use experimental measures of $N$ and $\epsilon$, seeming to be the
best constrained of the three, to estimate $\gamma$ and show that our
value of $\gamma$ is consistent with those experiments that have
measured ejecta velocities.

Here we consider the same target cases as \cite{farvokhar98} who
provide a very clear exposition of the effects of the Yarkovsky effect
on meter-class asteroids of various types. We assume our target
asteroids are either bare rock, regolith-covered rock or bare
iron. Bare or regolith-covered rock are the best studied in terms of
microcratering experiments, having received much attention during the
Apollo era \eg \cite{fecgauneu72}.

For the case of bare rock, perhaps the most likely situation for a
meteorite parent body, \cite{gau73} provides empirically-based
formulae for the displaced mass as a function of the kinetic
energy. After firing a variety of projectiles (densities of 0.95--7.8
g~cm$^{-3}$) into a selection of terrestrial rocks (including basalts)
as well as the Indarch meteorite (range of target densities: 2.5--5
g~cm$^{-3}$) at high-velocity and normal incidence, the mass displaced
$M_e$ in grams was found to be
\begin{equation}
M_e = 10^{-10.061} \left( \frac{\rho_p}{\rho_a} \right)^{1/2} \left(KE \right)^{1.133}
\end{equation}
where $\rho_p$ is the projectile density (g~cm$^{-3}$), $\rho_a$ is
the target density (g~cm$^{-3}$) and $KE$ is the kinetic energy in
ergs. Gault's formula is applicable to craters with diameters from
$10^{-3}$ to $10^{3}$~cm with impact kinetic energies of
$10-10^{12}$~ergs. An impact by a $1.5 \times 10^{-8}$~kg particle at
60 km/s has an energy of $2.7 \times 10^{8}$~ergs and falls squarely
in this range. The resulting $N=M_e/m$ is
\begin{equation}
N \approx 2.1 \times10^4 \left( \frac{\rho_p}{\rho_a} \right)^{1/2} \left( \frac{m}{1.5 \times 10^{-8}~{\rm kg}} \right)^{0.133} \left( \frac{v}{6 \times 10^4~{\rm m~s}^{-1}} \right)^{2.266} \label{eq:N}
\end{equation}
A hypervelocity impact can displace four orders of magnitude more
mass than that of the projectile.  For the case of dense minerals and
glass, microcratering is often accompanied by large spalled regions,
annuli around the crater itself where material fractures off in large
plates or flakes \citep[\eg][]{horhargau71} and this contributes to
the relatively large mass displaced.

The fraction of impactor kinetic energy that goes into the motion of
the ejecta has been found experimentally to be small for low impact
speeds but to increase sharply as speed increases. \cite{bra70} found
$\epsilon \approx 0.5$ both for dry quartz sand and basalt at impact
speeds of 6 km/s.  Later studies by \cite{har83} were somewhat
critical of Braslau's assumptions but still found $\epsilon \approx
0.3$ as the impact speed increased to 4~km/s. Here we will adopt $\epsilon =
0.5$ since we are considering even higher speeds, while noting that
our expression for $\alpha$ is relatively insensitive to its precise
value, going only like $\epsilon^{1/2}$ (Eq.~\ref{eq:alpha}).

Together with Eq.~\ref{eq:mult1} and \ref{eq:N} this allow us to
estimate $\gamma \approx \sqrt{\epsilon/N} \approx 5 \times
10^{-3}$. This value, which implies ejecta velocities around 0.3
km~s$^{-1}$ for a 60 km~s$^{-1}$ impact, is consistent with
experimental measurements of late-stage ejecta from loose sand targets
\citep{bra70} and powdered targets of pumice and basalt \citep{har85},
though the ejecta were found to have a wide range of velocities.

From the values for $\epsilon$ and $N$, we deduce an $\alpha \approx
\sqrt{\epsilon N} \approx 10^2$ (from Eq.~\ref{eq:alpha}) for bare
rock: the cratering process can release two orders of magnitude more
momentum than is carried by the projectile itself. As a result,
Eq.~\ref{eq:adot1} should be multiplied by a factor $\alpha \sim 100$,
which makes it competitive with the Yarkovsky effect under some
conditions (this will be discussed in more detail in section
\ref{extrapolation}).  We again note that our analysis is based on an
extrapolation of microcratering results beyond the impact speeds
actually examined in the lab, and a more detailed examination of the
phenomenon is certainly warranted.

Impacts into mineral dust, which would be more applicable to
regolith-covered bodies, are similar though the displaced masses are
often lower. High-velocity impacts by centimeter sized particles
displace two to four orders of magnitude more mass than that of the
projectile (references in \cite{ved72} incl. \cite{bra70}) with
loosely-packed material being more easily displaced than packed or
consolidated material.  For 2 to 5 micron-sized polystyrene (1.06~g
cm$^{-3}$) spheres accelerated to 2.5 to 12~km/sec into mineral dust,
the amount of displaced mass was three orders of magnitude larger than
the impactor mass, though the granular nature of the target resulted
in large ($\pm 50\%$) errors in displaced mass.  For impacts into
granular material, we conclude that $N$ may be lower, reducing
$\alpha$. Since small meteorite parent bodies are expected to be bare
rock, we will adopt $\alpha = 100$ here as our fiducial
value for stony targets, recognizing that it may be smaller if the
targets are actually covered with regolith.

For the bare iron case, \cite{com67} found that 2.6 km/s silicon
carbide particles eroded the Hoba iron meteorite at least two orders
of magnitude more slowly than the Indarch chondrite meteorite, implying
$N \las 200$ from simple extension of Eq.~\ref{eq:N}.
\cite{schnagneu79} fired $1.5-2$~mm steel and sapphire particles into
the Gibeon iron meteorite and also found that plastic flow
significantly reduced the amount of material released. They concluded
that the erosion rate of iron meteorites was a factor of 10 slower
than that of stony ones ($N\sim 2000$).

\cite{com67} points out that much depends here on the brittle versus
ductile nature of the target, which is sensitive to crystal size and
temperature. Higher velocity impacts at the lower temperatures of the
asteroid belt may move into the brittle fracture (instead of the
ductile flow) regime which may release more ejecta. However it seems
unlikely that $N$ values approaching those of rock targets would be
reached. If we adopt the values of $N\sim 200$ and take $\epsilon =
0.5$ as before, we have $\alpha = 10$ for iron meteorite parents,
recognizing it contains a substantial uncertainty. Nonetheless, even
on iron targets, it seems that significantly more momentum may be
carried by the ejecta than is received from the impinging particle.

If ejecta do generally carry away 10-100 times as much momentum as is
imparted by the meteoroid, then Eq.~\ref{eq:adot1} is 10-100 times
higher, and becomes competitive with the Yarkovsky effect. Before
comparing them side-by-side, let us first extrapolate the near-Earth
meteoroid environment to the asteroid belt.

\subsection{Extrapolation to the asteroid belt} \label{extrapolation}

Equation~\ref{eq:adot1} assumes a meteoroid environment like that at
Earth.  If we are interested in whether or not impact drag competes
with the Yarkovsky effect in the delivery of small asteroids to
resonance 'escape hatches' in the asteroid belt, we need to consider
the meteoroid environment there. Unfortunately, there is little
experimental data on the asteroid belt's meteoroid environment. Here
we simply assume that the environment at Earth has evolved directly
from that at the asteroid belt under PR drag.

The apex meteoroids encountered by our planet are on roughly circular
retrograde orbits \citep{jonbro93,chawoogal07}, and thought to be
particles released from retrograde comets whose semimajor axis $a$ and
eccentricity $e$ have decayed through Poynting-Robertson drag. We
assume here that the asteroid belt hosts the same meteoroids (at an
earlier time) as they spiral inwards to produce the meteoroids observed
at Earth, and that we can estimate the properties of the asteroid belt
meteoroid environment from meteor observations taken here.

Returning to Eq.~\ref{eq:adot0}, let's assume that the impactor mass
$m$, the target radius $R$ and density $\rho_a$ are not functions of
heliocentric distance. The factors that are include the velocity of
each impact $v$, which goes like $a^{-1/2}$ owing to Kepler's Third
Law. The flux of impactors $n$ is also affected: it is proportional to
$v$ times the number density of particles $n_p$. Assuming the
meteoroids are spiralling inwards under PR drag near the ecliptic
plane, the vertical and azimuthal densities of particles both go like
$a^{-1}$ while the radial density is proportional to $1/\dot{a}_{PR}$,
where $\dot{a}_{PR}$ is the rate at which the meteoroids spiral
inwards under PR drag. The effect of PR drag on particle semimajor
axis is given by \cite{weijac93} to be $\dot{a}_{PR} \propto a^{-1}$
for circular orbits. This result of these factors is that $n_p \propto
a^{-1}$, and the flux $n = v n_p \propto a^{-3/2}$. Eq.~\ref{eq:adot0}
would increase like $a^{3/2}$ if $n$ and $v$ were constant, but
$\dot{a}$ will instead decrease slightly (as $a^{-1/2}$) with
increasing semimajor axis.

When extended to the distance of the main asteroid belt,
Equation~\ref{eq:adot1} becomes
\begin{equation}
\dot{a} = -4.3 \times 10^{-4}  \left( \frac{s}{0.5} \right) \left( \frac{\alpha}{100} \right) \left( \frac{a}{2~{\rm AU}} \right)^{-1/2} \left( \frac{R}{\rm{1~m}} \right)^{-1} \left( \frac{\rho_a}{3500~\rm{kg~m}^{-3}} \right) ^{-1}    {\rm AU~Myr}^{-1}   \label{eq:adot2}
\end{equation}
Eq.~\ref{eq:adot2} is expressed at the same heliocentric semimajor
axis (2~AU) considered by \cite{farvokhar98} in their analysis of the
Yarkovsky effect, which allows us to compare our results directly with
theirs.

\begin{figure}
\includegraphics[width=15cm]{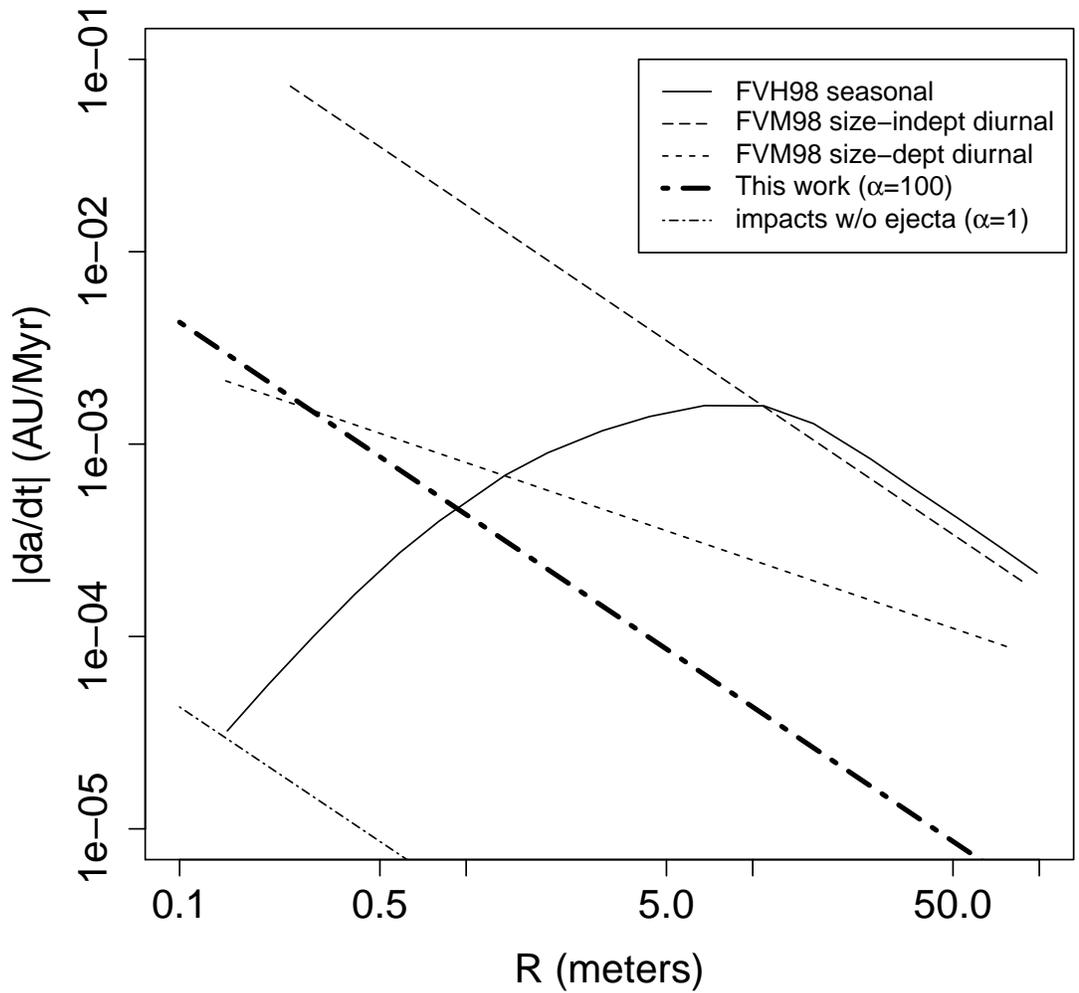}
\caption{The maximum semimajor axis drift for bare basalt fragment at
  2 AU. The curves for the seasonal and diurnal Yarkovsky effect,
  either with a size-independent spin period of 5h or with spin rate
  proportional to $1/R$ are adapted from \cite{farvokhar98} Figure 1. \label{fi:Figure1}}
\end{figure}

For bare rock (Fig.~\ref{fi:Figure1}) impact drag exceeds two of the
three variants of the Yarkovsky effect considered by
\cite{farvokhar98} (the seasonal effect and the diurnal effect under
the assumption of size-dependent spin) but only at sizes of tens of
centimeters.  The drag effect considered here is smaller than the
diurnal effect in the case where the spins of small asteroids are
independent of their sizes, though \cite{farvokhar98} indicate that
size dependent spin states are more realistic in their opinion. If
\cite{farvokhar98} are right and spin rates are typically higher for
smaller asteroids, then the effect presented here is of the same order
of magnitude as the Yarkovsky effect for meter class asteroids, while
the seasonal effect dominates at larger sizes.

\begin{figure}
\includegraphics[width=15cm]{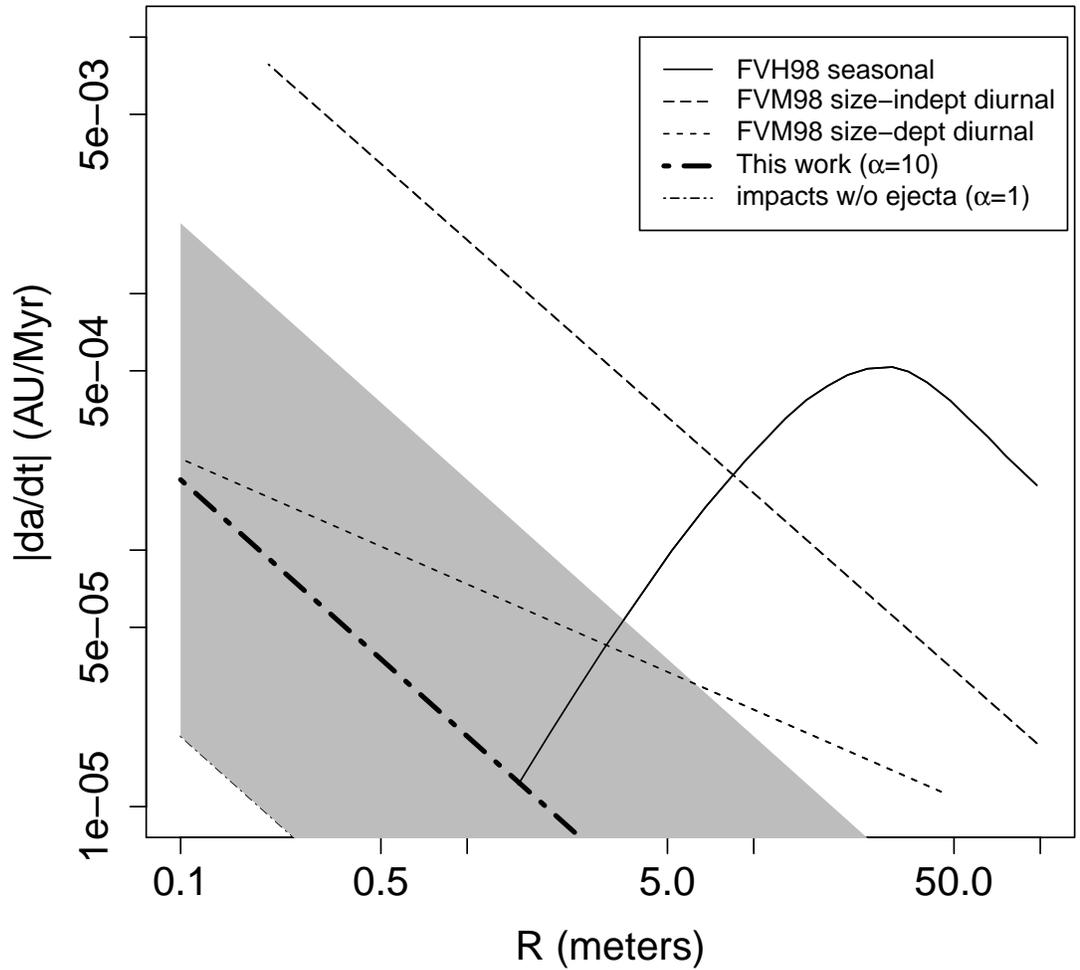}
\caption{The maximum semimajor axis drift for bare iron fragment at 2
  AU. The curves for the seasonal and diurnal Yarkovsky effect, either
  with a size-independent spin period of 5h or with spin rate
  proportional to $1/R$ are adapted from \cite{farvokhar98} Figure
  3. A grey region indicates a factor of 10 around the meteoroid drag
  line, indicating the high level of uncertainty in this value. \label{fi:Figure2}}
\end{figure}

For bare iron asteroids (Fig.~\ref{fi:Figure2}) meteoroid drag is less
effective, though it could still exceed the Yarkovsky effect at small
target sizes if 60km/s impacts take place in the brittle rather than
ductile deformation regimes (that is, if $\alpha$ is larger than
assumed here). In Fig.~\ref{fi:Figure2}, we account for the increased
density of the target ($\rho_a = 8000$~kg~m$^{-3}$ instead of $3500$) but
assume $\alpha = 10$. 

\begin{figure}
\includegraphics[width=15cm]{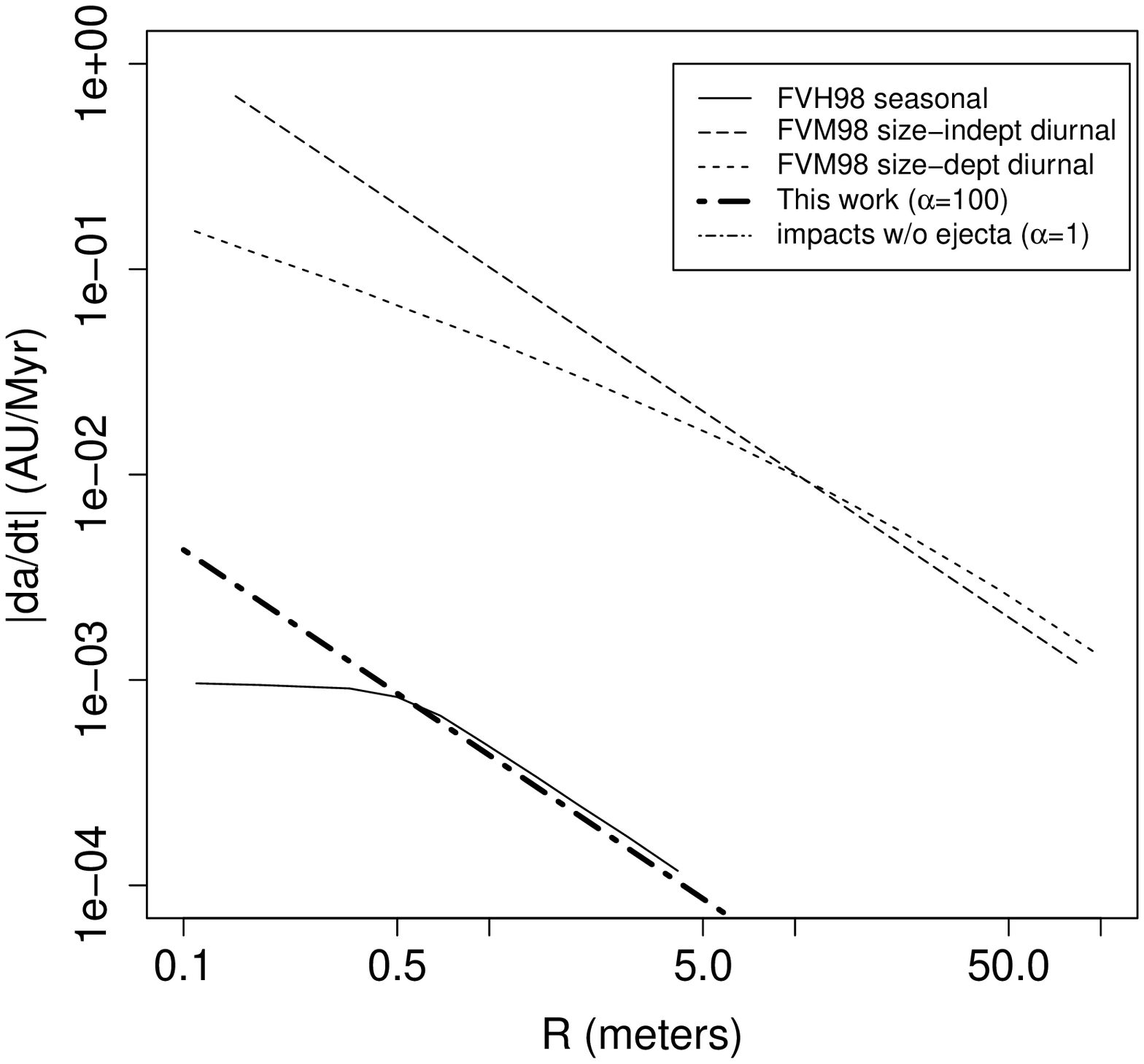}
\caption{The maximum semimajor axis drift for a regolith-covered
  fragment at 2 AU. The curves for the seasonal and diurnal Yarkovsky
  effect, either with a size-independent spin period of 5h or with
  spin rate proportional to $1/R$ are adapted from \cite{farvokhar98}
  Figure 4.\label{fi:Figure3}}
\end{figure}

For regolith covered rock (Fig.~\ref{fi:Figure3}), the drag effect
considered here falls far below the diurnal Yarkovsky effect, near the
seasonal.  The lower thermal conductivity of the regolith-covered body
enhances the Yarkovsky effect by increasing the day-night temperature
difference.

Since small asteroids tend to be more quickly rotating and have
smaller gravitational attractions, it has been argued that they are
unlikely to have substantial regolith \citep{farvokhar98}. If this is
correct and meter-class stony asteroids do not usually have regolith
coatings, then meteoroid impact drag will compete with Yarkovsky
effect in importance for stony meteorite delivery to Earth. It will be
less important in the delivery of regolith-covered bodies or iron
asteroids.

\subsection{Uncertainties}

Many unknowns also cloud the true importance of the effect of
meteoroid impact drag. Microcratering experiments seldom reach impact
speeds beyond 10 km/s due to the difficulty in accelerating the
projectile. The meteoroid impacts considered here are much faster:
does this increase or decrease the effective drag force? The meteoroid
environment at the asteroid belt is uncertain, and our estimate could
easily be off by an order of magnitude. Even the meteoroid flux at
Earth is not particularly well known. An earlier measurement of the
flux of impactors by \citep{gruzoofec85} had a number three times
smaller than that of \cite{lovbro93}. In the opposite direction, the
slope of the differential mass distribution of meteoroids near the
Earth is close to $-2$ \citep{blacamwer11,cambra11}. Values less than
$-2$ imply most of the mass is in the smallest particles (which is
what we've assumed here) while a value of precisely $-2$ means that
each decade of mass contributes equally. The proximity of the slope to
$-2$ means that our choice of typical impactor mass may be an
underestimate and may skew our projections downwards.

The physics of microcratering on an asteroidal target provide other
unknowns in terms of shape, composition and strength. For example,
impacts typically eject material on average perpendicular to the
normal of the surface except for the most oblique impacts
\citep{ved71}. This effect will lessen the back-reaction when the
impact takes place on the limb of the target.  On the other hand,
hypervelocity impactors are known to produce secondary craters
\citep{horhargau71}, which may themselves release more mass,
particularly if the impact takes place into a pre-existing concavity
such as a crater.  The breakage of edges and the release of
unconsolidated material, the rupture of the body or pieces thereof
\citep{gauhorhar72} may serve to increase the total momentum released
by an impact.

\subsection{Rotational state of the parent}

The two varieties of the Yarkovsky effect, the traditional or
'diurnal' \citep{opi51,pet76} and 'seasonal' \citep{rub95,rub98}
affect asteroid orbits differently. The diurnal effect can either
increase or decrease $a$ depending on the body's rotation state. The
seasonal effect always acts to decrease the semimajor axis, though its
magnitude also depends on the orientation of the asteroid's rotation
pole relative to the Sun. The impact drag effect considered here is
independent of the rotation state of the asteroid. However meteoroid
impacts can re-orient the spin of the asteroid and thus may play an
important indirect role in the Yarkovsky effect itself.

The order-of-magnitude assumption that is usually made in calculating
the time between pole reorientations involves equating the rotational
angular momentum of the asteroid to that imparted by the impactor to
determine the minimum size needed (at some typical encounter velocity)
to perform such a reorientation. A knowledge of the size distribution
of the impactors then allows the frequency of such impacts to be
estimated. However, the momentum carried by the ejecta also comes into
play here: the rotational momentum imparted by the impactor is not
just its own momentum times the radius of the target, but $\alpha$
times as much.  Though many poorly-understood effects (\eg asteroid
composition, shape, internal cohesiveness, the direction of debris
ejection under impacts near the limb of the asteroid, \etc) would come
into play in a detailed calculation, to first order the angular
momentum transferred to the asteroid is increased by a factor of
$\alpha$ over that usually assumed, which means that a particle of
only $\alpha^{-1}$ of the mass or $\alpha^{-1/3}$ the radius can
effect the same rotational change. For $\alpha \sim 100$, this
translates into a decrease by a factor of $100^{1/3} \approx 4.6$ in
radius. Given that the cumulative distribution of impactor sizes goes
something like $R^{-5/2}$ \citep{doh69} this translates to
reorientation events occurring $\alpha^{5/6} =100^{5/6} \approx 46$ times more
frequently.

Thus, the net effectiveness of the diurnal variant of the Yarkovsky
effect in particular may be considerably reduced. Meteoroid impacts do
not reduce the height of the Yarkovsky curves in Fig.~\ref{fi:Figure1}
- \ref{fi:Figure3}, instead they cause the sign of the effect to
change more frequently. If the distance between the asteroid and the
resonance is $\Delta a$, then the escape process will be of the nature
of a random walk if the time between pole reorientations $\tau_{rot}$
times $\dot{a}$ is greater than $\Delta a$. The more frequently
reorientations occur, the more asteroids will be in the random-walk
regime. Such asteroids will have a net Yarkovsky drift that proceeds
at a rate that is diminished by roughly one over square root of
$\alpha^{5/6}$ ($\sim 0.15 \sim 1/7$). This additional factor may mean
the smaller but consistently directed effect of meteoroid impacts can
compete with the diurnal Yarkovsky effect at even larger sizes than
Fig.~\ref{fi:Figure1} - \ref{fi:Figure3} would imply.

\section{The YORP effect} \label{YORP}
The Yarkovsky-O'Keefe-Radzievskii-Paddack (YORP) effect can change an
asteroid's spin through the uneven re-radiation of thermal
photons. Meteoroid impacts onto an asteroidal surface can change the
rotation rate of the target at rates which are below (but perhaps
uncomfortably close to) those currently being reported for YORP
detections in near-Earth asteroids.

The angular momentum transferred to the target in a single impact is
roughly $\Delta L = \alpha m v R$, while the target's initial angular
momentum is $L = I\omega$ where $I$ is its moment of inertia (here
taken to be that of a sphere $I=\frac{2}{5}MR^2$) and $\omega$ is its
angular rate of rotation, related to its period $P$ through $P=
2\pi/\omega$. The fractional change in angular momentum from a single
impact is
\begin{eqnarray}
 \frac{\Delta L}{L} &=& \frac{\alpha m v R}{\omega \left(
  \frac{2}{5} M R^2 \right)} \nonumber \\ &=& \frac{15 \alpha m v
  P}{16 \pi^2 \rho_a R^4} \label{eq:DeltaLL}
\end{eqnarray}
If we ignore the small change in the moment of inertia of the target
under the erosive effect of the impacts, then change in angular
momentum produces a concomitant change in the rotation rate of the
target, $\Delta \omega/\omega \approx \Delta L/L$. However, since
individual impacts occur randomly on the surface, they are as likely
to speed up the rotation as slow it down.

Let's consider the first reported detection of the YORP effect by
\cite{lowfitpra07,taymarvok07}. The asteroid (54509) 2000~PH5 was
observed to have its period decreasing at a fractional rate of $-1.7
\times 10^{-6}$ per year.  Assuming a mean radius of
57~m and a rotation period of 730 seconds for 2000~PH5,
Eq.~\ref{eq:DeltaLL} becomes
\begin{eqnarray}
\frac{\Delta \omega}{\omega} &\approx& 1.6 \times 10^{-10}  \left(\frac{\alpha}{100}
\right) \left( \frac{\rho_a}{3500~\rm{kg}~\rm{m}^3} \right)^{-1}\left(\frac{m}{1.5 \times 10^{-8}~\rm{kg}} \right)\left(\frac{v}{60~\rm{km}~\rm{s}^{-1}} \right) \label{eq:DeltaLL2}
\end{eqnarray}
so each impact affects the rotation rate by only one part in
$10^{10}$. Though minuscule, this is still much larger than the ratio
of the impactor to target masses, which is of order $10^{-17}$, and
hints that meteoroids may be more effective at changing asteroid
rotation rates than might initially be assumed. The difference arises
from the high speed of the impactor relative to the rotation of the
target: material even on the surface of 2000~PH5 moves no quicker than
about 0.5 m/s under rotation. The impactor contains $10^5$ times as much
momentum per unit mass as the target material, and momentum transport
by ejecta multiplies it further.

The asteroid suffers an impact rate $ s n \pi R^2$ which in this case
results in approximately $5000$ impacts per year on 2000~PH5. Assuming
a simple one-dimensional random walk, the net fractional change in the
period of order $10^{-8}$ per year. This is two orders of magnitude
below the fractional change in period observed ($-1.7
\times10^{-6}$~yr$^{-1}$) and so small apex meteoroids do not affect
the spin rate at the same level as YORP in this case.

A more recent determination of the YORP effect on asteroid (25143)
Itokawa \citep{lowweidud14} is also essentially unaffected by impacts
from apex meteoroids.  The rotation period is 12.14 hours and its mean
radius, 162~m \citep{schabeyos07}.  Despite its larger size, because
of its slower rotation, the fractional change in angular momentum per
impact is also $1.5 \times 10^{-11}$. Itokawa's larger size means that
the rate of impacts is slightly higher, approximately $4 \times
10^{4}$ per year, or a fractional change under a random walk of $3
\times 10^{-8}$~yr$^{-1}$. This is equivalent to a change in the
rotation period of $\sim 1$~millisecond over the course of one year,
much less than the value reported by \cite{lowweidud14} of $\sim
45$~ms~yr$^{-1}$.

The $R^{-4}$ dependence of Eq.~\ref{eq:DeltaLL} implies that the
rotation states of smaller asteroids are more susceptible to change by
meteoroid impacts.  If the random walk goes like the square root of
number of impacts ($\propto R^2$), then the net effect should go like
$R^{-3}$. This means that for the calculated effect of meteoroids of
1~ms~yr$^{-1}$ to increase to 45~ms~yr$^{-1}$, a decrease in the
asteroid size by only a factor of four is required. Thus the impact of
high-speed apex meteoroids may significantly influence the rotation
state (at least to the same degree as YORP) for asteroids smaller than
a few tens of meters in size.

\subsection{Single impacts}
The relatively large effect that a single meteoroid can have raises
the question of the smallest meteoroid impact that would produce a
result comparable to that of YORP. Here we will examine the single
impact required to create a fractional change in period of one part in
$10^6$ in 2000~PH5, comparable to that reported by \cite{lowfitpra07}
and \cite{taymarvok07} for the YORP effect.

Rearranging Eq.~\ref{eq:DeltaLL} gives a minimum impactor mass $m$
needed to produce a given fractional change in rotation rate
\begin{eqnarray}
m &=& \left( \frac{\Delta \omega}{\omega} \right) \frac{16 \pi^2 \rho_a R^4}{15 \alpha v P}\\
  && 8.9 \times 10^{-5} \rm{kg} \left( \frac{\Delta \omega/\omega}{10^{-6}}\right) \left(\frac{\alpha}{100} \right)^{-1} \left(\frac{\rho_a}{3500~\rm{kg}~\rm{m}^{-3}} \right) \left( \frac{v}{60 \rm{km/s}} \right)^{-1} \label{eq:deltaomega}
\end{eqnarray}

At 60 km/s impact speed, a fractional change in period of $10^{-6}$
could be generated by a single a 0.09 g meteoroid if $\alpha=100$. Such
a particle is approximately 2~mm in radius at a density of
2500~kg~m$^{-3}$. 

The apex sources are rich in small particles but not in large
ones. Though 2~mm radius particles certainly occur there, it is clear
that in considering the relatively large asteroids examined here,
impacts by larger particles will be more effective.  We instead consider
the effect of the sporadic meteoroid population as a whole, which has
somewhat lower speeds but more larger particles.

The encounter velocities between typical sporadic meteoroids and the
Earth are lower ($\sim 30$~km/s, \cite{cam08}) than for apex
meteoroids and so the impact speeds for asteroids on near-circular
prograde orbits near our planet are reduced as well. The excavated mass
rises in Eq.~\ref{eq:N} due to the larger projectile mass but
decreases by a comparable amount due to the smaller impact speed, so
we continue to adopt $\alpha=100$.  At this reduced speed, a mass of
0.18 g (Eq.~\ref{eq:deltaomega}) is required to affect a spin change
of one part in $10^6$ in 2000~PH5.

Using video recordings of meteors in Earth's atmosphere,
\cite{cambra11} found the total sporadic meteor flux to be $0.18\pm
0.04$~km$^{-2}$~hr$^{-1}$ ($5 \times 10^{-11}$~m$^{-2}$~s$^{-1}$) down
to a limiting mass of $2 \times 10^{-6}$~kg and deduced a differential
mass slope of $-2.02 \pm 0.02$. The cumulative size distribution is
then proportional to the impactor size to the -3.06, which is steeper
than the standard Dohnanyi -2.5 value: using the Dohnanyi value would
overestimate the impact rate somewhat.

A differential mass slope of near -2 implies that the cumulative
impact rate is inversely proportional to the mass, and so the rates of
impacts by meteoroids of at least $1.8 \times 10^{-4}$~kg is
approximately $(1.8 \times 10^{-4}/2 \times 10^{-6})^{-1} (5 \times
10^{-11}) = 5 \times 10^{-13}$~m$^{-2}$~hr$^{-1}$, or one every few
years on 2000~PH5, which can be neglected.

The change in period reported for Itokawa \citep{lowweidud14} is
similar, 45~ms in 12.14 hours or $\Delta P/P \sim 10^{-6}$. Though it
is a larger body, owing to its slower rotation rate, an impactor of
about the same mass as for the 2000~PH5 case is needed. Itokawa has a
larger cross section though, and might see one such impact per year
(assuming a mean radius of 162~m), a rate which is arguably
not entirely negligible.

Though meteoroid impacts may have an important role to play, we are
not asserting here that the spin changes attributed to YORP have
actually been produced by meteoroid impacts, for two reasons. First,
the observed changes in rotation are seen to be accelerating, which is
more compatible with YORP than impacts (though admittedly a small
number of impacts could conspire to look like a net acceleration in
the short term). Secondly, the analysis above uses a value of $\alpha$
of 100 deduced for bare rock. Itokawa at least is certainly
regolith-covered \citep[\eg][]{fujkawyeo06} which is likely to reduce $\alpha$. If the impact
rate is goes like the inverse of the impactor mass, then the rate of
impacts capable of generating the required change in period goes
roughly like $\alpha$. A small reduction in $\alpha$ would then reduce the
rate of impacts of concern to one every several years, which can be
neglected. Nonetheless, we do conclude that the effect of impacts is
at a level which demands some attention when sensitive measurements of
asteroid spins are being made.

\section{Other considerations} \label{misc}
\subsection{Other sporadic meteor sources and radiation pressure}
There are six generally accepted sporadic meteoroid sources, that is,
six broad inhomogeneities in the time-averaged meteoroid environment
seen by the moving Earth \citep[\eg][]{sto86, brojon95,chawoogal07,
  cam08}. The north and south apex sources have been the basis of the
analysis so far. Two others are the north and south toroidal sources,
which also arrive at the Earth from the direction roughly opposite its
motion but at higher ecliptic latitudes. These may contribute to the
drag force considered here but are generally weaker than the apex
sources and will be neglected here. The two remaining sources are the
helion and antihelion sources, consisting of particles on
high-eccentricity orbits which hit the Earth from the directions of
the Sun and of opposition respectively. These particles may create a
small radial force component on asteroids but have been ignored here
so far because the helion and antihelion sources have roughly equal
strengths, and so the net force from them will average out. Though it
is perhaps worth noting that the strengths are not precisely equal,
the antihelion source may in fact be stronger but probably only by
about 20-30\% (see \cite{wievaucam09} and references therein.)
 
One might wonder if the helion or antihelion meteoroid sources, owing
to the radial nature of the forces they create, could confound
measurements of the radiation pressure on small asteroids, which have
been used to determine their densities in particular cases
\citep{micthoell12, micthoell13, micthoell14}. We can show that the meteoroid
drag as considered here is much smaller than radiation pressure near
the Earth. Even if we ignore the opposing nature of the helion and
antihelion sources and assume that one or the other constitutes all of
the meteoroid impacts considered in Eq.~\ref{eq:accel}, the effect
is less than radiation pressure.  Adopting the expression of
\cite{burlamsot79} for the ratio of radiation pressure to solar
gravity and combining that with Eq.~\ref{eq:accel} yields a ratio of
radiation to meteoroid-derived accelerations $\beta_m$ of
\begin{equation}
\beta_m = \frac{L_{\odot} Q_{PR}}{4 \pi c s n m v r^2 \alpha} \sim 7 \times 10^{3}
\end{equation}
where $L_{\odot}$ is the solar luminosity, $Q_{PR}$ is a radiation
absorption coefficient we have taken to be unity, $c$ is the speed of
light, $m$ and $v$ are the impactor mass and velocities and $r$ is the
heliocentric distance. Here we have adopted $s=0.5$ and an impact
velocity $v$ = 30 km/s more appropriate for the helion and anti-helion
sources but this makes little difference; the effect of meteoroid drag
is much smaller than radiation pressure under all reasonable
conditions.

\subsection{Erosion rates}

The large amounts of ejecta produced by meteoroid impacts also
contribute to the erosion of the target body. The $1.5 \times
10^{-8}$~kg impactor considered here releases $3.1 \times 10^{-4}$~kg
of ejecta if $N=2 \times 10^4$. At the Earth's orbit this translates into a
rough survival time $\tau$ against erosion of a stony body
\begin{equation}
\tau \sim 16 {\rm Myr} \left( \frac{\rho_a}{3500 \rm{kg}~\rm{m}^{-3}} \right) \left( \frac{R}{1~\rm{m}} \right) \label{eq:tau}
\end{equation}
where this simple expression is an upper limit, as it ignores the
decreasing impact rate as the target size decreases.  This time is
comparable to the dynamical lifetimes (10~Myr,
\cite{glamigmor97,glamicfro00}) of near-Earth asteroids and so the
high levels of ejecta production assumed here do not conflict with
reality on this basis. The high erosion rates proposed here are also
consistent with cosmic ray exposure ages of meteorites. The cosmic ray
exposure ages of stony meteorites rarely exceed 100 Myr \citep{her05}
which can be accommodated in Eq.~\ref{eq:tau} by a parent body of ten
meters in size.



\section{Conclusions} \label{conclusions}

We have discussed the effect of the meteoroid environment
on small asteroids. It was argued that the net effect of such impacts
is enhanced by up to two orders of magnitude by the momentum transported
by ejecta. Careful examination of the physics of hypervelocity impacts
will be needed to determine the exact magnitude of the effect and its
broader role in asteroid evolution.

The instantaneous value of the net drag produced by the apex
meteoroids is found to be typically smaller than that of the variants
of Yarkovsky effect except at sizes well below one meter.  The effect
is independent of thermal or rotational properties of individual
asteroids, though not their densities. Independence from the
rotation state means that the meteoroid environment acts consistently
as a drag and cannot increase an asteroid's semimajor axis.  Impacts
also serve to reorient the spin axis of the target and thus can
decrease the net effectiveness of Yarkovsky drift.

The meteoroid impacts have the potential to confuse measurements of
the YORP effect. Here the effect is primarily due to larger
(centimeter) sized particles from the general sporadic meteoroid
population. However we conclude that impacts have probably not clouded
recent measurements of YORP among the near-Earth asteroid population,
though it will be an important consideration when measurements of
smaller (10 meter class) bodies are made. In fact, high precision
asteroid spin measurements may be sensitive enough to measure the
effect of meteoroid impacts on spin states.

The effects of meteoroid impacts on the dynamics of small asteroids
remains to be worked out in detail. If the momentum transport by
ejecta proposed here correctly represents real meteoroid-asteroid
collisions, then the meteoroid impacts may prove to be as important as
radiative effects in the dynamical evolution of small asteroids and
the transport of meteorites to Earth.

\section{Acknowledgements}
This work was performed in part with funding from the Natural Sciences and Engineering Research Council of Canada.

\section*{References}

\bibliography{Wiegert}

\end{document}